\documentclass[]{article}

\usepackage{amsmath}
\usepackage{amssymb}
\usepackage{mathtools}
\usepackage{tikz}
\usepackage{graphicx}
\usepackage{booktabs}
\usepackage{enumitem}
\usepackage{todonotes}
\usepackage{url}
\usepackage{cleveref}
\usepackage{authblk}
\usepackage[backend=bibtex]{biblatex} 
\title{A Classification Scheme for Local Energy Trading}
\author[1]{Jens H\"onen}
\author[1]{Johann L. Hurink}
\author[2,3]{Bert Zwart}
\affil[1]{Faculty of EEMCS, University of Twente, Enschede, The Netherlands}
\affil[2]{Department of Mathematics and Computer Science, Eindhoven University of Technology, Eindhoven, The Netherlands}
\affil[3]{Centrum Wiskunde \& Informatica (CWI), Amsterdam, The Netherlands}
\date{}

\begin{document}

\maketitle

\begin{abstract}
	The current trend towards more renewable and sustainable energy generation leads to an increased interest in new energy management systems and the concept of a smart grid. One important aspect of this is local energy trading, which is an extension of existing electricity markets by including prosumers, who are consumers also producing electricity. Prosumers having a surplus of energy may directly trade this surplus with other prosumers, which are currently in demand.\\
	
	In this paper, we present an overview of the literature in the area of local energy trading. In order to provide structure to the broad range of publications, we identify key characteristics, define the various settings, and cluster the considered literature along these characteristics. We identify three main research lines, each with a distinct setting and research question. We analyze and compare the settings, the used techniques, and the results and findings within each cluster and derive connections between the clusters. In addition, we identify important aspects, which up to now have to a large extent been neglected in the considered literature and highlight interesting research directions, and open problems for future work.\\
	\textbf{Keywords} Local energy trading  \and game theory \and distributed optimization \and smart grid \and literature review \and survey
\end{abstract}

\section{Introduction}\label{intro}

In the last few years, a growing awareness of the impact of climate change has led to a trend towards more sustainable and renewable energy sources, such as photovoltaic systems (PV) or wind power. This results in high penetration of renewable energy sources and can lead to severe problems within the energy system. Among the most crucial disadvantages are the lack of control on the generation side as well as the uncertainty in predicting future amounts of energy generation due to the intermittent nature of renewable energy sources. Especially within the electricity domain, this leads to an enormously complex, and challenging situation, as the ongoing electrification also leads to a drastic increase and changing patterns of electricity consumption.\\

In order to avoid upgrading the entire physical infrastructure of the electricity grid to compensate for the increased loads and peaks, one promising approach is the concept of a \textit{smart grid}. Its core is the addition of a two-way communication layer on top of the current physical grid and the introduction of advanced sensor technology and smart devices in the system. The keys to unlocking the potential of smart devices supporting the smart grid are algorithms and energy management systems. They use the additional information from the sensors, as well as information from outside the grid system, such as e.g. weather forecasts, and aim to use the potential advantage of the flexibility offered by smart devices. As such, corresponding control algorithms are an important tool in ensuring a stable electricity distribution in the future.\\

One core task of the new energy management system is to support and enable local energy trading. Historically seen, energy trading has always happened on national electricity markets, where electricity producers, grid operators, retailers and suppliers, as well as large energy-intensive companies buy and sell electricity and short-term capacities. Due to the ongoing penetration of newly affordable technologies such as PV systems, batteries, or electric vehicles (EVs) and in order to ensure a stable grid, it is essential to include also prosumers, which are consumers who also produce, store and trade energy by e.g. PV and battery systems, in future energy markets. Currently, prosumers are not part of energy trading yet, mainly because of their small size and the uncontrollable nature of their local energy production. However, due to new concepts such as microgrids (see \Cref{SModel}), these prosumers are now able to enter the energy markets. Possible approaches include trading with other prosumers, often referred to as \textit{Peer-to-Peer} trading (P2P), or acting as one larger unit on traditional energy markets by combining their flexibility. \\

The field of traditional energy markets has always attracted the attention of researchers from different areas, such as electrical engineering, operations research, or (power) economics. This research considered amongst others, variations of the classical unit commitment problem (e.g., \cite{UCPshorttermLR1988Bard}, \cite{AROUnitCommitment2013Bertsimasetal}, \cite{UCPequityefficiencyCompetitiveMarkets1997JohnsonOrenSvoboda}, \cite{UCPN-1SecurityWindUncertainty2019Sundaretal}), bidding or pricing strategies for different national electricity markets (e.g., \cite{EMentryexitgamesMeanFieldGame2021AidDumitrescuTankov}, \cite{OptTradingIntradayMarket2015AidGruetPham}, \cite{EMoptimalOffer2002AndersonPhilpott}, \cite{EMtransmissionPricing1997BushnellOren}, \cite{EMPriceModelAssetValuation2012CarmonaCoulonSchwarz}), load forecasting (e.g., \cite{ForecastingElectricLoad2009Hahnetal}, \cite{ProbabilisticForecasting2016Hongetal}, \cite{ForecastLongTerm2019Lindbergetal}), AC or DC optimal power flow computations (e.g., \cite{OPFchanceConstrainedUncertainty2012Bienstocketal}, \cite{OPFACmathematicalFormulations2020Bienstocketal}, \cite{OPFACstrongNPhard2019BienstockVerma}, \cite{OPFORguide2021SkolfieldEscobedo}) or cascading failures and blackouts in high-voltage grids (e.g., \cite{HVGsensitivityAnalysisCascadingFailures2012Bernsteinetal}, \cite{ETSCascadesOR2015BienstockB},  \cite{HVGMixedIntegerBlackout2007BienstockMattia},  \cite{FailuresCascadesPGridStatisticPhysic2018NestiZoccaZwart}). Compared to these problems, local energy trading focuses on the low-voltage (LV) or distribution grid and offers new, interesting challenges, such as congestions in LV grids due to the increased demand and local generation, or the impact of prosumers' behavior, which is not necessarily restricted to cost savings or profit maximization. Hence, existing techniques and approaches used on a national level cannot simply be transferred to the local level, but need to be adapted and newly implemented. \\

The main contribution of this work is twofold. In a first step, we provide a detailed overview of existing literature in the area of local energy trading. Due to the wide range of settings and research questions in publications in this field, we first identify and define various characteristics and research questions related to the settings. Based on these questions and characteristics, we classify the considered literature into three main clusters and analyze each cluster on its own. During this analysis, we compare the used techniques within each cluster, as well as investigate possible connections between the clusters. In a second step, we identify open questions and challenges in local energy trading, based on the insights gained from the analysis. To the best of our knowledge, many of these interesting open challenges have been neglected up to now.\\

The paper is structured as follows. In \Cref{SFindings}, we present our main findings and results, followed by a detailed classification scheme in \Cref{SClustering}. \Cref{SModel} introduces a general model of a microgrid, while \Cref{LitStudy} presents a short theoretical introduction of the techniques used in the considered literature. In \Cref{SAnalysis}, we analyze the different clusters in detail w.r.t. the settings and used techniques. We conclude the work with a summary as well as an outlook to future research directions and open problems within the scope of local energy trading in \Cref{SConclusion}.

\section{Findings and Results}\label{SFindings}

In this section, we give a summary of the three main findings identified within the scope of this work.
\begin{enumerate}
	\item \textbf{Three research directions:} Within the area of local energy trading, there are three main research questions, which up to now have been considered in the literature. The first research question deals with the problem of matching (fixed) demand and (fixed) supply. The second research question focuses on internal pricing schemes for the prosumers, in which each prosumer profits from participation. The last research question deals with the problem of a decentralized balancing of demand and supply for a set of prosumers, such as a microgrid. In contrast to the first research question, here smart devices and some parts of the demand and supply offer flexibility.
	\item \textbf{Grid constraints:} In contrast to classical energy markets, the considered literature rarely took grid constraints in the LV grid into account. While this may be reasonable and feasible for the classical energy production and generation of the last decades, the increased demands due to electrification as well as the local generation of e.g. PV systems require that (LV) grid constraints are also included in future local energy markets.
	\item \textbf{Uncertainty:} Only in a very limited subset of the literature uncertainty, either in prices or forecasts, is taken into account. The vast majority of the work assumes to have perfect knowledge of all parameters for these aspects. While this may seem reasonable for short time horizons, energy trading schemes that plan the trading for a whole day should include some form of safety against deviations from the predicted values.
\end{enumerate}
As a first conclusion, we can state that there are currently three main lines of research in the area of local energy trading. Herein, very interesting ideas and algorithms ranging from game theory to distributed optimization have already been applied, but some important aspects, in particular grid constraints and uncertainty, have been neglected or ignored to a large extent. Hence, the area of local energy trading still offers many interesting open problems for each of the given research lines.

\section{Clustering of Local Energy Trading Approaches}\label{SClustering}
Given the wide range of different settings and research questions in the considered literature, we first cluster similar problem settings together and then compare the clusters with each other. In order to cluster the settings, we first need to identify and define the key characteristics as well as the high-level research questions. Combining own findings as well as some of the characteristics found in \cite{MarketFrameworkLET2018KhorasanyMishraLedwich} and \cite{ClassificationElectMarket2005Ventosaetal}, the final classification scheme is based on the following characteristics:
\begin{enumerate}
	\item \textbf{Valuation}: Is the valuation of electricity of each prosumer taken into account? The valuation can be represented by means of bids in an auction approach or by utility functions in (non-)cooperative games. Utility functions are often a weighted sum of different aspects of electricity consumption, such as the cost of purchase, the profit of selling electricity, or the satisfaction of (the results of) electricity consumption.
	\item \textbf{Flexibility}: What kind of flexibility w.r.t. electricity consumption and generation is taken into account? This can span from no flexibility over flexibility due to the usage of a battery or EV to flexible load and curtailment of PV generation. 
	\item \textbf{Structure}: How is the computation organized? Is it done centrally at the microgrid operator (MGO) or are computations distributed among all participants?
	\item \textbf{Objective}: What is the main objective of the setting? The objectives in the considered literature range from maximizing social welfare, over minimizing costs to minimizing peak load or maximizing local consumption.
	\item \textbf{Stability and Fairness}: Should the solution be stable against strategic manipulation of participants? Is the solution fair to everyone? As there is no common scientific definition of fairness, it refers to the lack of discrimination of a subset of the prosumers within assignments or decisions taken by the algorithms. 
	\item \textbf{Participants and their incentives}: Which entities can participate in the local energy trading scheme? Do all participants profit from the local energy trading scheme or may some participants be off the same as when not joining the energy trading scheme?
\end{enumerate}

Using these characteristics, the final classification scheme consists of three major clusters, each with a unique setting and high-level research question. Two of the major clusters can be further split into two subclusters each. The reason for maintaining such a structure with three main clusters and two of them consisting of two subclusters each is that the two subclusters are very similar to each other, meaning that they are much closer related to each other than to any other cluster. Hence, we keep the three main clusters, each with a clear focus on the setting and problem definition, but also acknowledge the smaller but still noticeable differences within two of the clusters. We refer to \Cref{fig:2} for an overview of the relation between the different clusters and the considered characteristics.\\

The first identified cluster consists of settings in which the valuation of electricity of each prosumer is taken into account in the form of a bid. Demand and supply are fixed and smart devices do not offer any flexibility. Only prosumers are considered participants in this cluster and no participant is worse off compared to not joining the proposed energy trading schemes. The high-level research question in this setting is to match fixed demand and supply with each other while maximizing the social welfare of the participants. In the following, we refer to this cluster as the \textit{Matching without Flexibility} cluster. A closer analysis of the settings in this cluster reveals that it can be split further into two subclusters. In the first subcluster, all computations are done centrally and, assuming rational participants, the solution is protected against strategic manipulations. In addition, not all participants need to profit from participating. This subcluster is referred to as the \textit{Strategic Matching without Flexibility} cluster. In the second subcluster, on the other hand, computations may be done in a decentralized way, but there is no guarantee that the solution is stable. Given some (light) assumptions, all participants may profit from their participation. We refer to this subcluster as the \textit{Direct Matching without Flexibility} cluster. \\

The second cluster does not take the valuation of prosumers of electricity into account. Instead, its goal is to minimize the overall costs of a given set of prosumers and divide the cost among the participants. Demand and supply are fixed, but in some cases, flexibility is offered by batteries. The computation is done centrally, and the distribution of the cost should be done fairly and in such a manner that no group of participants has an incentive to deviate from the centrally managed solution. The high-level research question is to find a pricing scheme, such that every prosumer benefits from following the optimal solution. For the remainder of this work, we refer to this second cluster as the \textit{Cooperative Pricing Scheme} cluster.\\

The third cluster encompasses settings in which flexibility is offered by smart devices, such as batteries, EVs, or heat pumps, but also by flexible parts of the load. Participants can include a wide range of entities, from prosumers to companies or aggregators. The valuation of electricity is usually taken into account in the form of a utility function, often as a weighted sum of different objectives. Computations are done in a decentralized way, usually mimicking the underlying structure of the participants. The focus of the high-level research question is to locally balance consumption and generation using the given flexibility while maximizing social welfare. In the following, we refer to this cluster as the \textit{Balancing with Flexibility} cluster. Similarly to the \textit{Matching without Flexibility} cluster, we again can split the cluster into two subclusters, each with its own focus. In the first subcluster, the whole group of participants shares a common objective, usually the sum of the utility functions of all participants. The main goal is then to find an optimal solution that maximizes (or minimizes) the objective. We refer to this subcluster as the \textit{Joint Balancing with Flexibility} cluster. The other subcluster on the other hand treats every single participant as an individual, selfish agent which wants to maximize its own utility. In this setting, the objective is to find an equilibrium solution in which no participant can improve its utility by deviating from this solution. In the following, we refer to this cluster as the \textit{Equilibrium Balancing with Flexibility} cluster.\\

\begin{table}
	\centering
	\begin{tabular}{l | c | c | c | c | c|}
		& \multicolumn{2}{c|}{MF} & & \multicolumn{2}{c|}{BF}\\
		Characteristics & SMF & DMF & CPS & JBF & EBF \\ 
		\midrule
		Valuation: &  &  &  &  & \\
		- bids & x & x & - & - & -\\
		- utility function & - & - & - & - & x\\
		\midrule
		Flexibility: &  &  &  &  & \\
		- battery & - & - & x & x & x\\
		- demand & - & - & - & x & x\\
		- supply & - & - & x & x & x\\
		\midrule
		Structure: &  &  &  & & \\
		- centralized & x & x & x & - & -\\
		- decentralized & - & - & - & x & x\\
		- hybrid & - & x & - & x & x\\
		\midrule
		Objective: &  &  &  &  & \\
		- max social welfare & x & x & - & - & x\\
		- min total cost & - & - & x & x & -\\
		\midrule
		Stability and Fairness: &  &  &  &  & \\
		- stability & x & - & x & - & x\\
		- fairness & - & - & x & - & -\\
		\midrule
		Participants and Incentives: &  &  &  &  & \\
		- participants & all & all & all & all & all\\
		- incentives & x & x & x & - & x\\
	\end{tabular}
	\caption{Overview of the relation between the different clusters and the characteristics; Matching without Flexibility (FM), Balancing with Flexibility (BF), Smart Matching without Flexibility (SMF), Direct Matching without Flexibility (DMF), Cooperative Pricing Scheme (CPS), Joint Balancing with Flexibility (JBF), Equilibrium Balancing with Flexibility (EBF))}
	\label{fig:2}       
\end{table}

At the start of our research, we considered in addition to characteristics 1 to 6 also the following characteristics for the classification scheme. However, the additional characteristics were not selected because either they did not add any further insights into the clusters, or the resulting (sub)clusters were too small and the differences between subclusters were only present in one single aspect.
\begin{enumerate}
	\setcounter{enumi}{6}
	\item \textbf{Devices:} What kind of (smart) devices, such as PV systems, batteries, EVs, or heat pumps are considered in the approach? \\
	Within the scope of local energy trading, we are much more interested in the flexibility the devices can offer compared to the management of the device themselves. Hence, the actual devices are not important, but only the (type of) flexibility they can offer.	
	\item \textbf{Time Horizon:} What is the considered time horizon of the setting? Are multiple time slots considered at once or only one single time slot after the other?\\
	The characteristic of time horizon is only important in the presence of smart devices, such as batteries or EVs, which can shift energy demand through time. Within each of the three main clusters, either all approaches have the same time horizon, or approaches covering one time slot do not consider the flexibility of such smart devices, and therefore each time slot can be optimized individually.	
	\item \textbf{Grid Constraints:} Are grid constraints considered in the problem definition?\\
	Due to an increase in electrification of mobility and heating as well as local electricity generation, congestion in LV grids or other violations of (LV) grid constraints are becoming more likely and hence pose a serious threat to the reliability of future electricity distribution.\\
	However, grid constraints are rarely taken into account in the considered approaches. Hence, the resulting subclusters, which do take grid constraints into account, only consist of one or two approaches, and therefore can rather be seen as outliers than as actual clusters which offer further insights into the underlying structure of local energy trading.	
	\item \textbf{Uncertainty:} Is uncertainty taken into account?\\
	In particular in settings with a larger time horizon, forecasts and predictions of load and generation are often not perfect. PV generation heavily depends on the weather, while the household load is subject to the prosumer's decisions and behavior. Both, human behavior, as well as the intermittent generation of renewable energy sources, are known to be difficult to predict.\\
	While it is reasonable to assume that in settings with only one time slot, uncertainty does not play a large role due to the short time horizon, most settings with larger time horizons also do not consider uncertainty in predictions and forecasts. Similar to the grid constraints, the resulting subclusters would be very small and not yield any additional insights into local energy trading.
\end{enumerate}

\section{General Model}\label{SModel}
Before analyzing the settings and approaches within each cluster, we provide a common ground w.r.t. the used models and techniques. We first give a brief outline of the current electricity system, followed by a generalized framework, which includes the time horizon, a model of a microgrid as well as possible ways of trading. Although the model is quite comprehensive, specific aspects, such as e.g. the incorporation of time-shiftable devices, are not taken into account. Nevertheless, all considered publications can be described well enough using the given model and trading possibilities.\\

\textbf{Electricity System:} The current electricity system, as can be found in Europe, can be split up into two layers. The top level is the high-voltage or transmission grid. Large energy generators, such as e.g. coal- or gas-fired power plants, as well as large-scale consumers, such as the energy-intensive industry, are directly connected to this level. The \textit{transmission system operator} (TSO) is responsible for the safe and reliable transmission of electricity between the producers and consumers on this level and the mid/low-voltage grid. The bottom layer consists of the mid/low-voltage or distribution grid which connects the high-voltage grid with electricity consumers, like private households or smaller businesses. The \textit{distribution system operator} (DSO) is responsible for the safe and reliable operation on this level. Note, that details regarding the distribution of responsibilities, here in the form of the TSO and DSO, may differ from country to country. Local energy trading mostly takes place in the mid- and low-voltage grid.\\

\textbf{Time Horizon:} In general, energy management approaches cover a specific time horizon, which is discretized into equally-length, non-overlapping time slots $t \in \mathcal{T}$, where $\mathcal{T}$ denotes the set of all time slots and $t_1 < t_2$ indicates that time slot $t_1$ ends before time slot $t_2$ starts. Based on this discretization, load and generation can be represented by vectors of length $\vert \mathcal{T} \vert $, with the $t-$th entry representing the load, respectively the generation, in time slot $t$. Depending on the chosen setting, the time horizon spans a certain time period, which is often one day or a part of a day, although the period may also be longer. In literature often a time slot covers one hour.\\

\textbf{Prosumer:} A microgrid represents a small part of a distribution grid, consisting of a set of prosumers $\mathcal{N} = \left\lbrace 1, \ldots, N \right\rbrace $. To specify the electricity behavior of a prosumer $i\in \mathcal{N}$, devices and parameters may be used, which define the main characteristics of the corresponding household of the prosumer:
\begin{itemize}
	\item The uncontrolled or fixed load $fl_i \in \mathbb{R}^{\vert \mathcal{T} \vert }_{\geq 0}$ of the household over the considered time horizon, whereby $fl_i^t$ represents the fixed load of the household resulting during time slot $t$. This load results from devices that are assumed to not be shiftable in time without a massive change in human behavior, such as e.g. TV, lightning, or cooking.
	\item The adjustable load $al_i \in \mathbb{R}^{\vert \mathcal{T} \vert }_{\geq 0}$ of household $i$, which can be adapted by reducing (curtailing) it, implying that only an amount between $0$ and $al_i^t$ has to be served in time slot $t$. Note, that $al_i$ represents the maximum amount of flexible load added to the fixed household load.
	\item A battery, which can be used to increase the demand by charging or acting as a source of electricity by discharging. It has charging and discharging limits, a capacity limit, and an initial state of charge at the beginning of the given time horizon. Batteries are especially of interest for prosumers also producing electricity. In literature, various battery models are known and can represent different levels of detail, such as charging and discharging losses.
	\item An electric vehicle (EV), from a modeling perspective, can also be seen as a battery with some additional restrictions, resulting from arrival and departure times and a required state of charge at the departure time. In literature often settings are considered, where an EV is not discharging and thereby not providing electricity to the prosumer.
	\item Devices, which are a source of electricity, such as e.g. PV panels. Comparable to the adjustable load, the PV generation of a household $i$ during time slot $t$ can be specified by a value $p_i^t \geq 0$. The generation may be curtailable, implying that the delivered electricity then has to be between $0$ and $p_i^t$. 
	\item An individual goal of household $i$, such as e.g. minimizing energy costs, maximizing the self-usage of PV generation, or maximizing the own utility of consuming energy. This individual goal is modeled as a utility function $u_i$, which may reflect a combination of different goals and maps an energy profile of the household to a value. This value then represents the valuation of household $i$ for the profile and the utility function can be used to evaluate and compare different profiles with each other. Hereby, the utility function may depend, next to the profile vector of prosumer $i$, also on the energy profiles, and thereby the decisions, of other prosumers.
\end{itemize} 

\textbf{Microgrid Operator:} Next to the households, also a microgrid operator (MGO), which is responsible for the interaction of the microgrid with the main grid, is a relevant entity in a microgrid. In some research, this MGO also has control over certain assets, such as e.g. a battery. In this case, the battery serves as a sort of communal battery and can be used to improve the overall situation for the microgrid. The MGO has also its own utility function $u_{MGO}$, which may represent different goals of the community, such as e.g. minimizing the interaction with the main grid, minimizing costs of the microgrid, maximizing self-usage of PV generation, or minimizing peaks in the energy profile of the overall microgrid. Additionally, some papers consider a setting, where the MGO is also responsible for the compliance of low-voltage grid restrictions and limitations.\\

\textbf{Trade:} A last relevant aspect of a microgrid is its connection to the main grid or the market. For this, mainly the cumulative difference between the load and generation of the whole microgrid is of interest. This difference has to be served by or exported to the main grid. In most literature, it is assumed that the corresponding prices are fixed, whereby the feed-in price is in general assumed to be strictly smaller than the price for buying electricity from the main grid.\\

Interactions take place between the introduced entities of a microgrid, e.g. an interaction between two prosumers represents a trade between them. In the literature, mainly the following two types of trades are considered:
\begin{itemize}
	\item \textit{Direct} trade between prosumers, where a prosumer $i$ sells electricity to a prosumer $j$. This type of trade is often referred to as \textit{peer-to-peer} (P2P) trading. Note, that in most cases the overall demand in a microgrid is not equal to the overall supply, and therefore, also the possibility to trade with the main grid needs to be considered.
	\item \textit{Indirect} trade between prosumers, where prosumers do not directly trade with each other, but with the MGO. They either sell or buy electricity to/from the MGO, which thereby serves as an intermediate agent linking demand and supply within the microgrid. In order to balance demand and supply within the microgrid, the MGO can interact with the main grid, and thereby participate in e.g. day-ahead markets or trade with traditional suppliers. For this type of trading also the term \textit{community trading} is used.
\end{itemize}

\begin{figure}
	\centering
	\includegraphics[width=0.75\textwidth]{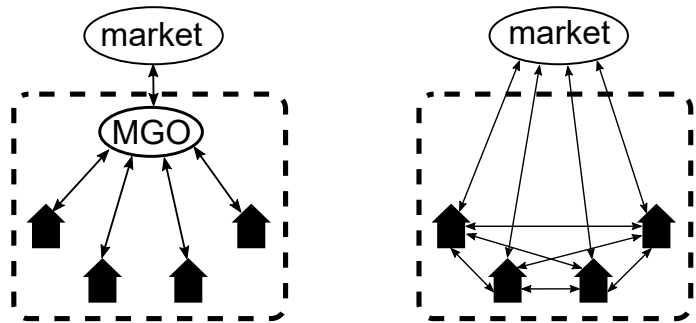}
	\caption{Sketch of trading relationships between prosumers and the market in indirect and direct trading}
	\label{fig:1}       
\end{figure}

See also \Cref{fig:1} for a simplified representation of indirect and direct trading. The houses within the dotted rectangle represent the prosumers of the microgrid, which communicate either only with the MGO in the indirect trading form or directly with each other in the direct trading. Note, that there exist various hybrid forms of community trading and P2P trading, such as e.g. neighboring microgrids trading with each other and with traditional markets, see e.g. \cite{P2PReview2019Sousaetal}. In addition, it should be mentioned that concepts similar to local energy trading are also designed for structures such as energy communities, in which households are not necessarily geographically close to each other.

\section{Theoretical Background}\label{LitStudy}
Based on the general model presented in \Cref{SModel}, many ways to trade between different entities with the goal to balance demand and supply are possible. Some of the approaches leave the decisions completely to the prosumers, while others use signals to steer the prosumers in a certain direction. In this section, we introduce and explain general concepts and theories behind the different approaches presented in the considered literature. These approaches cover techniques from game theory over distributed optimization techniques to economic theories. For a more in-depth analysis of the techniques and proofs, we provide additional references. Readers who are familiar with techniques and concepts such as e.g. the core, the Shapley value, Stackelberg games, or ADMM may continue with the detailed analysis in \Cref{SAnalysis}.\\

The research field of game theory deals with mathematical models and concepts for the strategic behavior of rational players. A common division of game theory is into the areas of cooperative and non-cooperative game theory. As the name already hints, cooperative game theory deals with aspects related to coalitions of players, such as the allocation of profit of cooperation among all participating members. Often, the main focus is on finding such an allocation that makes the coalition of all players stable, meaning that no subset of members has an incentive to leave the coalition. Non-cooperative game theory, on the other hand, focuses on the behavior of players who compete with each other. Concepts such as an equilibrium, which is a situation where no player can be better off by deviating, are the main foundation for determining allocations. A further important area, which lies in the intersection of game theory and optimization is mechanism design, which focuses on designing market rules to ensure socially desirable outcomes. These concepts are explained in more detail in the following three subsections, followed by a fourth subsection, where relevant decentralized optimization techniques for the energy area are sketched.\\

\subsection{Cooperative Game Theory}
Starting with techniques from cooperative game theory, we introduce the concepts of the core and the Shapley value. Given a coalition game $\Gamma = (N,v)$, with $N$ being the set of players, in our case the prosumers, and $v:2^N \rightarrow \mathbb{R}$ being the value function, assigning a value $v(S)$ to each subset of players $S \subseteq N$, the main question of cooperative game theory is how to allocate the value of a coalition to its members. In the area of energy trading, the value function $v(S)$ could correspond to the value of the energy savings of a coalition $S$ given that prosumers in $S$ cooperate with each other e.g. by trading or using batteries. Let $x \in \mathbb{R}^{\vert N \vert}$ denote an allocation with $x_i$ being the share of player $i\in N$. An allocation $x$ is called \textit{feasible}, if $\sum_{i \in N}x_i \leq v(N)$, that is the allocation distributes not more than the value of the grand coalition $N$. The allocation can be seen as the distribution of the overall savings of all prosumers among themselves.\\

The Shapley value is one of the most well-known concepts in cooperative game theory. Its main goal is to characterize a \textit{fair} allocation of value to the players. It is based on the following three axioms:
\begin{itemize}
	\item \textbf{Symmetry}: For all $i, j \in N$, where $v(S\cup \left\lbrace i\right\rbrace ) = v(S \cup \left\lbrace j\right\rbrace )$ for all $S \subset N$ with $i,j \notin S$, the allocations are equal, that is $x_i = x_j$.
	\item \textbf{Dummy player}: A player is a dummy player if she always adds the same amount of value to any coalition she joins, i.e. for all $S$ with $i \notin S$ we have $v(S \cup \left\lbrace i\right\rbrace ) - v(S) = v(\left\lbrace i\right\rbrace )$. The allocation for a dummy player is then $x_i \coloneqq v(\left\lbrace i\right\rbrace )$. This axiom is sometimes also known as the \textit{null player} axiom.
	\item \textbf{Additivity}: For two games $\Gamma_1=(N,v)$, $\Gamma_2=(N,w)$ over the same set of players $N$, and the game $\Gamma_3=(N,(v+w))$, defined by $(v+w)(S) \coloneqq v(S)+w(S)$ for all $S \subseteq N$, the allocation for each player for game $\Gamma_3$ has to be equal to the sum of the allocations for the two games $\Gamma_1$ and $\Gamma_2$, i.e. we have $x_i((v+w)) = x_i(v) + x_i(w)$.
\end{itemize}
In \cite{AGTBalancedSetsCores1967Shapley}, Shapley specified an allocation scheme satisfying all these three axioms. In addition, it was shown that this is the unique allocation $x$ satisfying efficiency, that is $\sum_{i \in N}x_i = v(N)$. The Shapley value of player $i$ for a game $\Gamma = (N,v)$ is given by
\begin{equation*}
x_i \coloneqq \dfrac{1}{\vert N \vert !} \sum_{S \subseteq N \setminus \left\lbrace i\right\rbrace } \vert S \vert ! (\vert N \vert - \vert S \vert -1)! \left[ v(S \cup i) - v(S) \right] .
\end{equation*}
It can be seen as the average marginal contribution of player $i$ to any coalition. Taking a closer look, we see that in order to compute the Shapley value using this formula we need to consider every subset at least once. Therefore for general coalition games, the running time to compute the value in this way is exponential in the number of players. Nevertheless, there are approaches to deal with the running time, for example by approximating the value using (random) subsets of the coalitions of the players. Furthermore, for some special value functions, there also exists closed-form formulations that can be computed efficiently.\\

While the Shapley value is seen as fair and always exists for any coalition game, the question arises whether this allocation is also \textit{stable}. The stability of an allocation is related to the question of whether subsets of agents could get better off by forming smaller coalitions on their own. One concept that deals with the stability of an allocation is the core, which is the set of stable allocations. An allocation $x$ is stable and therefore in the core of a game $\Gamma=(N,v)$, if and only if 
\begin{equation*}
\sum_{i \in S}x_i \geq v(S), \quad \forall S \subseteq N.
\end{equation*}
In practical applications, this is a desirable property, as a stable allocation ensures that no subset of players would have an incentive to deviate from the grand coalition. One important question is whether there always exists an allocation in the core. The answer to this is 'no', meaning there are games with an empty core, implying that there is no allocation to the players that is stable. For games with a non-empty core, the next interesting problem is if the Shapley value always lies within the core. Once again the answer to that question is 'no', as there are games where the core is non-empty, but the Shapley value is not part of it. In \cite{EoGT2008LeytonBrownYoav} some examples of such games are given, as well as a detailed introduction to the area of cooperative game theory.\\

\subsection{Non-cooperative Game Theory}
In this section, we consider the research field of non-cooperative game theory. A (finite) normal-form game is defined by a triple $\Gamma = (N, \mathcal{A}, u)$, with $N$ being a (finite) set of players $i$, $\mathcal{A}= \mathcal{A}_1 \times \mathcal{A}_2 \times \ldots \mathcal{A}_n$ the set of available actions, with $A_i$ being the action space of player $i$, and $u:\mathcal{A} \rightarrow \mathbb{R}^n$ the utility function, mapping a feasible strategy $s \in \mathcal{A}$ to its value $u(s)$. Hereby, the $i$-th entry of $u(\cdot)$ represents the utility of player $i$. In the context of local energy trading, the set of players may represent the set of prosumers of the microgrid, and the action space $A_i$ of player $i$ may represent the possible actions and decisions of prosumer $i$ during a given time horizon $\mathcal{T}$. Often this translates to energy schedules specifying for each time slot $t$ the usage of energy due to devices such as an EV or batteries. The utility function $u_i(s)$ in the game represents the utility function of prosumer $i$, as mentioned in \Cref{SModel}. In most cases, the utility function is a combination of the cost of energy given the profile $s$ and the preferences of a prosumer w.r.t. its energy usage. There are two solution strategies for the players: 
\begin{itemize}
	\item \textbf{Pure strategy}: Each player $i$ chooses a pure action $a_i \in \mathcal{A}_i$.
	\item \textbf{Mixed strategy}: Each player $i$ randomizes over her set of available actions following some probability distribution. A mixed strategy of player $i$ is specified by a vector $s_i \in \mathbb{R}^{\vert \mathcal{A}_i \vert}_{\geq 0}$ with $\sum_{a_i \in \mathcal{A}_i}s_i(a_i)=1$ and $s_i(a_i)$ the probability that action $a_i$ is chosen by player $i$. We denote the space of all mixed strategies of player $i$ by $S_i$.
\end{itemize}
Note that a pure strategy is also a mixed strategy with all probability on one action. In contrast to the above definition, in the energy trading context, a mixed strategy can be seen as a convex combination of multiple (pure) strategies rather than a probability distribution, as each pure strategy corresponds to an energy usage profile that can be combined with each other. In a general game $\Gamma$, this may not be possible and therefore the terms of mixed and pure strategies are used. The (expected) utility of a strategy $s$ for player $i$ is given by
\begin{equation*}
u_i(s) = \sum_{a \in \mathcal{A}} u_i(a) \prod_{j=1}^{n}s_j(a_j).
\end{equation*}
If the game has only 1 player, this player can directly decide whether a strategy is optimal or not. However, in general, player sets are larger and we need another way to evaluate a strategy. To deal with this, we introduce one of the arguably most important solution concepts in game theory, namely the \textit{Nash equilibrium} (NE). A strategy profile $s=(s_1,s_2,\ldots,s_n)$ is called a NE, if and only if for all players $i$
\begin{equation*}
u_i(s_i, s_{-i}) \geq u_i(s_i^*, s_{-i}) \quad \forall s_i^* \in S_i,
\end{equation*}
where $s_{-i}$ is the strategy profile $s$ without the strategy of player $i$. Referring back to the situation of energy trading, strategy profile $s$ is a NE, if and only if no prosumer $i$ can improve her utility (function) by deviating from $s_i$ given that all other prosumers still act according to $s_{-i}$. In \cite{AGTNashEquilibriumNGames1950Nash}, Nash introduced his idea and showed that for every finite $n$-player game, there exists a mixed NE.\\

One popular game used for modeling markets is the \textit{Stackelberg game}. It is often used for oligopoly models, where one player moves first, and the other players can observe this move and then decide on their actions. The player moving first is called the \textit{leader} of the game, and the remaining players are the \textit{followers}. Depending on the exact setting of a Stackelberg game there either exists a closed-form solution or an iterative approach converging towards the Stackelberg (Nash) equilibrium (SNE). In the context of energy trading or energy management systems, the leader is often the MGO, or another third entity independent of the prosumers, such as a DSO. The utility function of this leader is then related to grid constraints, such as minimizing peaks and its action space consists of setting energy prices for the prosumers. The followers are the prosumers, who, based on the decision of the MGO, try to maximize their own utility. They then report their resulting power profile back to the MGO, which can either react by updating the prices (increasing prices during times of peaks) and thereby entering the next iteration or by accepting the actions of the prosumers. The resulting solution is an SNE that ensures that neither the prosumers nor the MGO can be better off by deviating from the chosen actions.\\

\subsection{Mechanism Design}
The research area of (algorithmic) mechanism design lies within the intersection of game theory and optimization. Its goal is to design algorithms or rules that ensure a socially desirable outcome in settings with selfish decision-making agents having individual preferences. Examples of such settings include various auctions, voting systems, cost-sharing mechanisms, or matchings.\\

The difference between a mechanism design and an optimization viewpoint is that in mechanism design the agents cannot directly be forced to reveal their preferences. In addition, agents may try to manipulate the outcome by revealing false preferences, and they may not have an incentive to accept the outcome. The main challenge now is to design the algorithm such that the individual objectives of the agents align with the overall goal of the socially desirable outcome. Often, the algorithm can be split up into two parts, namely the assignment and the payment scheme. The goal of the assignment scheme is to efficiently find a socially desirable solution assuming that the agents reveal their true valuation. The payment or incentive scheme then needs to be designed in such a way that revealing the true preferences always results in the best outcome for each agent.\\

In the following, we use the example of a \textit{second-price sealed bid} auction to highlight how assignment and payment schemes can look like. We also shortly introduce other types of auctions without going into the details of the algorithms.\\

In the \textit{second-price sealed bid} auction, an auctioneer wants to sell only one item. The set of agents is the set of the possible buyers, each with a private valuation of the item. The utility function of an agent $i$ is either 0 if agent $i$ does not get the item after the auction, or it is its valuation minus the price paid for the item. The socially desirable outcome for the auctioneer is to sell the item to the agent with the highest valuation. During the auction process, each agent can submit a value, representing its public valuation, without the remaining agents knowing the value. Note, that the bid is not necessarily equal to the private valuation. The auctioneer then declares the agent with the highest bid the winner (assignment scheme) and charges as the price the second highest bid to the winner (payment scheme). It can be shown that for each agent, truthfully submitting its private valuation is a dominant strategy. This property is often referred to as \textit{dominant-strategy incentive compatible} (DSIC). Thus, agents cannot improve their utility by misreporting their valuation, and due to the structure of the payments, they also have an incentive to accept the outcome.\\

The \textit{Vickrey-Clarke-Groves} (VCG) mechanism is a generalization of the above auction setting and is able to select any socially desirable solution out of a set of feasible solutions while being truthful. It can also be applied to other auction setting such as the \textit{double auction}, which is a two-sided auction. In two-sided auctions, the set of agents can be split into two subsets, namely the sellers and the buyers. Bids now consist of two values, the first one is the number or amount of items, in our case, electricity, which they either want to buy or sell. The second value is the price they are willing to pay or the price they want to receive. McAfee, \cite{DoubleAuction1992McAfee}, proposed a truthful mechanism for two-sided auctions which ensures that the payments received by the sellers are equal to the payments made by the buyers, a property which the VCG mechanism applied to this setting is not able to guarantee. Other variants of two-sided auctions are \textit{continuous double auctions}, in which bids arrive over time and after each arrival, the auctioneer checks for possible trades. For an in-depth review of different truthful mechanisms as well as a theoretical introduction to mechanism design, see \cite{AlgGameTheory2007NisanRoughgardenTardosVazirani}, in particular chapters 9 to 16.\\

In the area of energy trading, two-sided auctions, such as (continuous) double auctions have been applied for matching demand and supply. The players are the prosumers with either a surplus (seller) or a demand (buyer) of energy, as well as a private valuation of energy. The solution is then a set of players who trade with each other instead of selling to or buying from the electricity market. \\

Another approach from mechanism design, which is used for energy trading is called \textit{cake-cutting}. The problem is to divide an infinitely divisible set of heterogeneous resources among a set of players. The main difference to other allocation problems in game theory, such as the core or Shapley value, is that the players have individual value functions $u_i$ over the set of resources. As players are selfish, the task is to divide the resources such that the assignment is seen as fair by every player. There are several notions that are related to the term fairness, but for cake-cutting, a mechanism to divide the resources is \textit{proportional}, if for each player its piece is at least as valuable as the value of the complete resource divided by the number of players. A mechanism is said to be \textit{envy-free} if for each player its own piece is at least as valuable as the piece of any other player. Both proportionality and envy-freeness are popular properties of fairness, see \cite{DivConqPropMinEnvyCakeCutting2011BramsJonesKlamler}. In the area of energy trading, cake-cutting games have been used to split up the budget of an MGO for buying energy from its prosumers using different prices, see \cite{AGTPriceDiscriminationEnergyTrading2017Tusharetal}.\\

\subsection{(Decentralized) Optimization Techniques}
Apart from game theoretic techniques, there are also several optimization techniques, which have been successfully applied to problems in energy management and trading. Within this section, we focus on decentralized optimization techniques, but also provide an overview of popular centralized optimization techniques.\\

One popular decentralized technique is the \textit{alternating direction method of multipliers} (ADMM), which solves convex optimization problems by decomposing the problem into subproblems and solving these sequentially. The algorithm is applicable to optimization problems of the form
\begin{align*}
\text{min } & f(x) + g(z)\\
\text{s.t. }& Ax+Bz=c,
\end{align*}
where $x$ and $z$ are vectors of variables, $A$, $B$ and $c$ given matrices or vectors, and $f$ and $g$ given convex functions. The augmented Lagrangian of this problem is
\begin{equation*}
L_{\rho}(x,z,y) = f(x) + g(z) + y^T(Ax+Bz-c) +(\rho /2)\Vert Ax+Bz-c \Vert_2^2, 
\end{equation*}
with $\rho >0$. This function is minimized in an iterative approach, where, instead of optimizing over $x$ and $z$ at once, these vectors are updated one after the other, leading to the term \textit{alternating direction}. The corresponding iterative steps are
\begin{align*}
x^{k+1} := \text{argmin}_x L_{\rho}(x,z^k,y^k),\\
z^{k+1} := \text{argmin}_z L_{\rho}(x^{k+1},z,y^k),\\
y^{k+1} := y^k + \rho \left( Ax^{k+1} + Bz^{k+1}-c \right).
\end{align*}
Regarding convergence rates, there are many different results in the literature. One very general assumption, which is often made, is that the functions $f$ and $g$ are closed, proper, and convex. This assumption implies that the iterative subproblems of determining $x^{k+1}$ and $z^{k+1}$ are solvable. In practice, ADMM often converges fast to a moderate level of accuracy, but afterward shows slow converging behavior to high accuracy. See \cite{distOptStatLearningADMM2011Boydetal} for a detailed and in-depth analysis of ADMM, including both, theoretical and practical results.\\

A further class of decentralized optimization techniques, which have been successfully used in the energy domain, are variants of \textit{Consensus and Innovation} (C+I) \cite{P2PmarketProdDiffCandI2019SorinBoboPinson} or decentralized versions of primal-dual gradient methods \cite{P2PdecentralizedBilatTrading2020KhorasanyMishraLedwich}. These decentralized optimization techniques are applied by modeling the trading system using a mathematical formulation such as an LP, MILP, or MIQP. The arguably most important step here is to formulate the problem in such a way, that the formulation can be decomposed into subproblems. These subproblems hereby often represent the problem of energy scheduling for a single prosumer, which then can solve its own (sub)problem using standard optimization techniques, such as interior point methods or the Simplex algorithm. Several of the above-mentioned techniques rely on a master problem which coordinates the process of solving the overall problem. This master problem is often solved by the MGO and together with the subproblems being solved by the prosumers, this is a quite natural representation of the structure of the underlying microgrid. These approaches also show similarities to the Stackelberg games, in which the prosumers solve their own problems and communicate their solutions to the MGO, which in turn updates and sends prices or other steering signals.\\

As briefly mentioned, the resulting subproblems often still need to be solved using centralized optimization methods. Usually, it is not stated which exact technique is used, but depending on the format of the optimization problem, standard optimization approaches, such as interior point methods, the (dual) Simplex algorithm, or gradient descend methods, may be used. A heuristic optimization technique, which has been used is genetic algorithms.

\section{Analysis}\label{SAnalysis}
Based on the introduced general framework and the different techniques, we now study the introduced classification scheme in detail. We first describe the settings within each cluster and then analyze and compare the different techniques used within each cluster.

\subsection{Matching without Flexibility}

\subsubsection{Strategic Matching without Flexibility}
As shortly described in \Cref{SClustering}, the Strategic Matching without Flexibility cluster is one of two clusters that mainly focuses on matching prosumers to each other. Even beyond the considered characteristics 1 to 6, the settings are all very similar to each other. If devices, such as PV or batteries, are considered, they are usually not controlled within the scope of the used techniques, but rather change the demand or surplus of the corresponding prosumers in a fixed and often greedy way, without offering any further flexibility. Corresponding to this general setting of fixed demand and supply, and no flexibility, the considered time horizon usually only covers one time slot. Hence, the research questions aim to match prosumers with each other on rather short notice, such as given in balancing markets. This also aligns with the absence of uncertainty in the data. If the considered time slot is rather short and the computations are done directly before the realization, forecast errors may be reasonably small and can therefore be neglected. Analyzing the approaches within this cluster, it becomes obvious that also the used techniques are quite similar and often related to concepts from mechanism design. The techniques can be divided into three groups, namely auctions, non-cooperative games as well as approaches based on matching and contract theory.\\

Approaches based on auctions, are presented in \cite{MDMarketMechanismMicroCHP2008Blocketal}, \cite{DRtoTE2017ChenLiu}, \cite{P2PdecentralizedNetworkConstraints2018GuerreroChapmanVerbic}, \cite{TEAuctionTradingActiveProsumer2017KhorasanyMishraLedwich}, \cite{P2PMarketClearingKnapsack2017KhorasanyMishraLedwich}, \cite{HeatPowerDoubleAuction2019SaurYorkeSmithLaPoutre}, \cite{LocalMatchingFlexibility2016StroehleFlath} and \cite{P2PEnergyTradingGrid2019Tusharetal}. In all of these approaches, the prosumers participate in an auction and can be divided into two groups, one with a surplus of and one with a demand for electricity. The outcome of the auction is a subset of the prosumers, which trade with each other, as well as a clearing price and the amount of electricity each prosumer in the subset contributes to the trade. Different types of auctions have been proposed throughout literature, with the standard double-auction (\cite{TEAuctionTradingActiveProsumer2017KhorasanyMishraLedwich}, \cite{HeatPowerDoubleAuction2019SaurYorkeSmithLaPoutre} and \cite{P2PEnergyTradingGrid2019Tusharetal}) or combinatorial (VCG) auctions (\cite{DRtoTE2017ChenLiu}, \cite{P2PMarketClearingKnapsack2017KhorasanyMishraLedwich}) as presented in \Cref{LitStudy} being the most prominent ones. A very interesting way how to combine heat and electricity into one auction is presented in \cite{HeatPowerDoubleAuction2019SaurYorkeSmithLaPoutre}, where the double auction is modified to be run once for both energy types simultaneously. \cite{TEAuctionTradingActiveProsumer2017KhorasanyMishraLedwich} presents a way how to include grid constraints into an auction setting by modifying the pricing mechanism of the double auction to include additional charges based on a linear approximation of active power flow. In \cite{LocalMatchingFlexibility2016StroehleFlath}, the concept of a double auction is used in an online setting, in which both, demand and supply offers may appear and disappear over time. The approach in \cite{LETcontractGames2018Zhangetal} is based on the same principles as the auctions above, namely individual rationality (IR) and incentive compatibility (IC), but makes use of contract theory. All sellers publicly announce their producer type, which contains the amount of electricity to sell and the cost of production. The buyer then uses a mathematical model to find the optimal bids for each type. It is shown that for sellers truthfully reporting their type is a best response strategy. \\

Different matching approaches are presented in \cite{P2PEMPeerMatching2021Khorasanyetal} and \cite{AGTP2PPVtrading2021KrayemAhmadNajem}. In \cite{P2PEMPeerMatching2021Khorasanyetal}, the matching is done via a priority list which is based on the economic profit of a trade between two prosumers. The negotiation between two matched prosumers is executed as an iterative approach and the final solution is shown to be a NE. In \cite{AGTP2PPVtrading2021KrayemAhmadNajem}, the matching is based on the outcome of the Galey-Shapley algorithm, where the input is a distance measure of the difference between surplus and demand between each pair of prosumers. In this approach, there is no negotiation process as the internal trading price is fixed to a certain fraction of the trading price with the external grid. \\

In contrast to the previous approaches, the approach in \cite{AGTcontributionbasedEnergyTrading2016Parketal} is based on a non-cooperative game. A set of prosumers reports their demand or supply for the coming time slot to the MGO. Based on previous contributions to the microgrid, as well as its current request, the MGO distributes the surplus of electricity to the prosumers with a need. The distribution algorithm is based on a water-filling algorithm, while the strategic choice of how much energy to ask for is decided using a non-cooperative game among the prosumers with a demand for electricity.

\subsubsection{Direct Matching with Flexibility}

Another way to enable trading between prosumers is to either use continuous double auctions (\cite{MDMarketMechanismMicroCHP2008Blocketal}, \cite{P2PContinuousDoubleAuctionPredictionModel2019ChenLinSong}, \cite{P2PdecentralizedNetworkConstraints2018GuerreroChapmanVerbic}, \cite{TradingAgentSmartGrid2010Vytelingumetal}), or to match buyers and sellers directly with each other (\cite{P2PEMPeerMatching2021Khorasanyetal}, \cite{EnergyStorageinLEM2017MengelkampGarttnerWeinhardt}). Using the iterative nature of a continuous double auction, in \cite{MDMarketMechanismMicroCHP2008Blocketal} heat and electricity are traded together. To achieve this, two continuous double auctions are executed in parallel and after each new computation, bundle constraints between heat and electricity are checked. In \cite{P2PdecentralizedNetworkConstraints2018GuerreroChapmanVerbic} and \cite{TradingAgentSmartGrid2010Vytelingumetal}, each possible trade is first checked w.r.t. grid violations, before it is allowed and congestion prices may be added. In \cite{P2PEMPeerMatching2021Khorasanyetal} on the other hand, the peer matching algorithm includes the impact of trades for the grid constraints by means of the pricing negotiation. \cite{EnergyStorageinLEM2017MengelkampGarttnerWeinhardt} analyses the efficiency of random peer matching on the social welfare. The price negotiation can range from pay-as-bid strategies for the buyers (\cite{EnergyStorageinLEM2017MengelkampGarttnerWeinhardt}) to iterative negotiation algorithms (\cite{P2PEMPeerMatching2021Khorasanyetal}), in which both players update their prices until a final price is found. In \cite{LEMCompMD2017Mengelkampetal}, a comparison between continuous double auctions and random peer-matching algorithms with different price negotiation techniques is presented.\\

\subsection{Cooperative Pricing Scheme}

As mentioned in \Cref{SClustering}, the main research question for the settings in the Cooperative Pricing Scheme cluster is to encourage prosumers to follow the centrally computed solution by creating a pricing scheme from which every prosumer profits. The settings within this cluster are once again very similar to each other, also beyond the considered characteristics. Apart from \cite{AGTCoalitionalGameDirectElectricityTrade2014Leeetal}, no approach considers uncertainty in any form, although there are some settings with multiple time slots. Grid constraints are also not taken into account. This may be explained with the settings, where prosumers do not have flexibility in their load to change their demand or surplus. Hence, even without the centrally computed solution, prosumers would still trade the same amount of electricity with the grid. Therefore, no additional problems w.r.t. grid congestion or violation of grid constraints appear. In all settings, some form of renewable energy production, mostly PV generation, is considered. Some of the settings also include batteries, which are usually used to minimize the amount of traded electricity with the grid. The settings with batteries often cover multiple time slots, while most settings with only one time slot do not consider batteries. The main difference between the approaches, which consider only a single time slot at once, and approaches, which consider multiple time slots can be found in the centralized computation, while the pricing scheme is often identical or at least very similar. In general, the used techniques are based on cooperative game theory, in particular the Shapley value and the core.\\

The approaches, which are directly based on the Shapley value are \cite{AGTflexCooperativeGTP2pEnergyTrading2018Dixonetal}, \cite{P2PcooperativeGameKMeansCluster2019Hanetal}, \cite{AGTProsumerCoalitionsEnergySaving2018HanMorstynMcculloch}, \cite{AGTincentivizingCoalitionsEnergyManagement2019HanMorstynMcCulloch}, \cite{P2PShapleyEstimateStratSampling2021HanMorstynMcCulloch},  \cite{AGTCoalitionalGameDirectElectricityTrade2014Leeetal}, \cite{MultiMicrogridDSM2021Li}, \cite{AGTP2PenergyTrading2019LongZhouWu}, \cite{MarketMechanismHeatElectricty2021MitridatiKazempourPinson} and \cite{AGTbidrectionalPaymentCore2021Yamamoto}. While \cite{AGTflexCooperativeGTP2pEnergyTrading2018Dixonetal},  \cite{AGTCoalitionalGameDirectElectricityTrade2014Leeetal} and \cite{AGTbidrectionalPaymentCore2021Yamamoto} cover one only time slot, the remaining approaches solve the centralized problem of minimizing the total cost of the microgrid for a larger time horizon. Combined with batteries, this can further increase the cost saving compared to solutions of individual prosumers. One important aspect to be considered when using the Shapley value is the scalability of the approach. In general, an exponential number of subproblems has to be solved to be able to compute the Shapley value. Hence, for slightly larger microgrids this may already pose a serious problem. Fortunately, only the payments depend on the Shapley value, while the distribution of electricity between the prosumers and devices is the solution to the centralized optimization problem. Hence, the Shapley value may theoretically still be computed even after the considered time horizon. One assumption in this context is that all prosumers will participate in the trading scheme, even without knowing their exact cost savings. Another way how to avoid the problem of computational complexity is by approximating the Shapley value, which can be done in various ways, see \cite{P2PcooperativeGameKMeansCluster2019Hanetal}, \cite{P2PShapleyEstimateStratSampling2021HanMorstynMcCulloch} and \cite{AGTCoalitionalGameDirectElectricityTrade2014Leeetal}. In \cite{P2PShapleyEstimateStratSampling2021HanMorstynMcCulloch}, a stratified sampling approach is used to reduce the number of subproblems to solve, while \cite{P2PcooperativeGameKMeansCluster2019Hanetal} clusters similar prosumer profiles together to reduce the number of participants and thereby also the number of subproblems to solve. The approach in \cite{AGTCoalitionalGameDirectElectricityTrade2014Leeetal} does not reduce the number of subproblems to solve, but rather computes the asymptotic Shapley value using statistical parameters of the considered uncertainty. In addition, it is shown that the asymptotic Shapley value lies within the core of the cooperative game.\\

Another technique used within the Cooperative Pricing Scheme cluster is to show that a tailor-made pricing scheme lies within the core of the cooperative game, and is therefore stable against group deviations. As mentioned before, this is done in \cite{AGTCoalitionalGameDirectElectricityTrade2014Leeetal} for the Shapley value, but also in \cite{MarketMechanismHeatElectricty2021MitridatiKazempourPinson} for a pricing scheme based on the nucleolus and in \cite{AGTP2PEnergyTradeUserParticipation2018Tusharetal} for the mid-market price. The mid-market pricing scheme simply computes the internal trading price as the average of buying and feed-in price of the grid. Based on the common assumption that the feed-in price is strictly smaller than the buying price of the grid, it can be shown that this pricing scheme lies within the core of the game. Apart from the nucleolus and the Shapley value as pricing schemes, in \cite{MarketMechanismHeatElectricty2021MitridatiKazempourPinson} two additional allocation schemes are introduced. The first one is a uniform pricing scheme, while the second one is based on the VCG payment rule. In \cite{P2PEnergyTradingCommunityGrid2017Longetal} several pricing schemes are proposed. The first pricing scheme is the mid-market price, which lies within the core, as explained above. The second pricing scheme is based on a double auction, while the third one is referred to as bill-sharing. In this pricing scheme, the single microgrid bill of the overall trade with the external grid is shared among all prosumers using a fixed internal price for buying and selling.\\

A further technique from cooperative game theory is used in the second step of the two-step optimization approach in \cite{P2PEnergyTrading2StageOpt2021JiangYuanLi}. Here, a Nash Bargaining approach is used to decide how the jointly generated cost savings should be distributed among the players.

\subsection{Balancing with Flexibility}
\subsubsection{Joint Balancing with Flexibility}

The settings within the Joint Balancing with Flexibility cluster deal with the question of how prosumers can make use of their flexibility to optimize a common objective. The settings and approaches within this cluster are all rather similar to each other, although some settings highlight particular aspects, such as the preference between different 'types' of electricity, such as e.g. locally produced or green electricity. Due to the decentralized structure of computation, individual data and parameters of the participants, such as the flexibility or the valuation of electricity, do not need to be shared. Therefore, most settings take data privacy into account. As mentioned in \Cref{SClustering}, all of the settings in the Joint Balancing with Flexibility cluster do take flexibility into account. In some settings, this flexibility directly stems from devices such as batteries, EVs, or heat pumps, but there are also some problem definitions, in which flexibility only stems from the flexible part of the load. These settings often reduce the problem to a bare minimum and do not model any devices. Nevertheless, the models are able to represent the key problems that may occur in (local) energy trading. Regarding the time horizon, some settings directly formulate models for multiple time slots, while quite a few of the considered settings within this cluster only formulate single time slot models. Nevertheless, it is often noted that for the sake of simplicity and notation only a single time slot is modeled, but the presented approach can easily be adapted for multiple time slots. However, even though larger time horizons can be modeled at once, uncertainty is not taken into account in any of the settings. Grid constraints are also not considered by the majority of settings, although a few use approximations of power flow to create price signals for overloaded lines within the grid. In all settings, decentralized optimization techniques, such as ADMM, relaxed C+I, or decentralized primal-dual algorithms are used.\\

The approaches in this cluster are \cite{ProsumerMarketsUnifiedForm2019BarocheMoretPinson}, \cite{P2PRealTimeOnlineOptimization2021Guoetal},  \cite{P2PEnergyTrading2StageOpt2021JiangYuanLi}, \cite{DistMarketClearingLET2018KhorasanyMishraLedwich}, \cite{P2PdecentralizedBilatTrading2020KhorasanyMishraLedwich}, \cite{ConsensusUserPreferences2019LeCadreetal},  \cite{P2PAlgorithmsComputation2018Moretetal}, \cite{RiskPrefElecMarkets2020MoretPinsonPapakonstantinou}, \cite{P2PEnergyTradingMultiClassEM2018MorstynMcculloch}, \cite{P2PmarketProdDiffCandI2019SorinBoboPinson} and \cite{P2PReview2019Sousaetal}. In \cite{ProsumerMarketsUnifiedForm2019BarocheMoretPinson}, \cite{P2PRealTimeOnlineOptimization2021Guoetal}, \cite{ConsensusUserPreferences2019LeCadreetal}, \cite{P2PAlgorithmsComputation2018Moretetal}, \cite{RiskPrefElecMarkets2020MoretPinsonPapakonstantinou},  \cite{P2PmarketProdDiffCandI2019SorinBoboPinson} and \cite{P2PReview2019Sousaetal}, the approaches are based on simplified and reduced models in which no devices are directly modeled. Nevertheless, different types of local trading, such as direct peer-to-peer trading, communal trading, or a hybrid version are formulated and solved either via ADMM or relaxed C+I, see \Cref{LitStudy} for a short introduction to these techniques. Due to the structure of the simplified models, there is no difference in the objective value of the proposed decentralized and centralized optimization algorithms. In addition to an optimal solution, the relaxed C+I in \cite{P2PmarketProdDiffCandI2019SorinBoboPinson} also computes prices for each individual trade, which are based on the economic concept of shadow prices, that are the dual variables of the trade constraints. In \cite{DistMarketClearingLET2018KhorasanyMishraLedwich} and \cite{P2PdecentralizedBilatTrading2020KhorasanyMishraLedwich}, the same simplified model is extended by grid constraints in the form of distribution load flow. Based on the load flow, the power transfer distribution factor (PTDF), which computes the contribution of each trade between prosumers in the power flow, is computed for each line in the grid and is used as a price signal for the prosumers. Before solving this model with an adapted decentralized primal-dual gradient method, Lagrangian multipliers move global constraints into the objective function. Instead of introducing grid constraints to the simplified models, in \cite{P2PEnergyTrading2StageOpt2021JiangYuanLi}, \cite{P2PoptStrategyCommunity2019Leeetal} and \cite{P2PEnergyTradingMultiClassEM2018MorstynMcculloch} different devices, such as batteries are directly modeled. While \cite{P2PEnergyTradingMultiClassEM2018MorstynMcculloch} introduces prosumer preferences over different classes of electricity, such as green or local electricity, \cite{P2PEnergyTrading2StageOpt2021JiangYuanLi} focuses on a payment scheme in a second stage, which is based on a Nash Bargaining game. All three models are again solved using ADMM. In \cite{RiskPrefElecMarkets2020MoretPinsonPapakonstantinou} risk levels for prosumers are introduced to model different human behavior in the presence of uncertainty. The model is again solved using ADMM.

\subsubsection{Equilibrium Balancing with Flexibility}

Another approach to make use of flexibility is offered by the settings in the Equilibrium Balancing with Flexibility cluster. In contrast to the Joint Balancing with Flexibility cluster, the participants behave more selfishly and do not simply act as distributed computing units for the goal of the whole microgrid. Individual objectives and goals are more important and techniques that lead to stable solutions in which no participant can improve anymore have to be used. Beyond the considered characteristics, for most aspects, there are large similarities between the settings, although there are some exceptions. Comparable with the previous cluster, due to the decentralized structure of computations, in most cases, sensitive data, such as flexibility or utility functions, can remain private for each participant. Apart from \cite{GNEanalysisP2PMarketDistGrid2021ShilovLeCadreBuvsic}, no other setting considers grid constraints in its approach. Devices are mostly explicitly modeled, although there are a few settings, in which there are either no devices modeled, or PV generation is indirectly included via the load profiles. Regarding the considered time horizon, the settings are evenly split up between considering only a single time slot and multiple time slots at once. Furthermore, unlike the previous cluster, settings covering only one time slot can not always easily be upgraded to multiple time slot models. This is mainly a consequence of the absence of one central model which can be split up into subproblems for each participant. Adapting all individual models while ensuring that the used techniques still converge to an equilibrium is more challenging. Regarding uncertainty, only some settings take that into account, even if a larger time horizon is modeled. Hence, no exact pattern between time horizon and uncertainty can be recognized, as there are settings with only one time slot, but also settings covering multiple time slots, which consider uncertainty. Due to the focus on individual objectives, techniques in this cluster have to be able to represent this selfish behavior, while ensuring that a stable solution is found. Game theory offers the right tools for such problems, and in most settings, a Stackelberg game is used to model the relation between the different participants. Other settings ignore the leader-follower dynamic of Stackelberg games and focus on general non-cooperative games. In some settings, either the non-cooperative or Stackelberg games are complemented by other techniques, such as auctions.\\

The first group consists of Stackelberg games in which prosumers are leaders and followers. The notion of the prosumer is generalized beyond the definition in \Cref{SModel}, as also companies that either buy or sell electricity are included. This setting is considered in \cite{P2PLETVirtualMicrogrid2020Anohetal}, \cite{LETPriceUncertStackelGame2019ElRahietal}, \cite{DistETStackelGameEquilibrium2015Leeetal}, \cite{EnergySharingStochProgStackGame2018Liuetal}, \cite{P2PpricebasedDRMicrogrids2017Liuetal} and  \cite{P2PEnergyTradingProsumerCommMicrogrid2019Paudeletal}. In \cite{P2PLETVirtualMicrogrid2020Anohetal}, \cite{DistETStackelGameEquilibrium2015Leeetal} and \cite{P2PEnergyTradingProsumerCommMicrogrid2019Paudeletal}, the set of prosumers is divided into a set of sellers and a set of buyers. The sellers act as the leaders in a multi-leader multi-follower Stackelberg game, while the buyers are the followers. The strategies of sellers and buyers can differ from one approach to the other. In \cite{P2PLETVirtualMicrogrid2020Anohetal} and \cite{DistETStackelGameEquilibrium2015Leeetal}, the sellers start by announcing the amount of electricity they are willing to sell, and the buyers react with the prices they are able to pay. Based on these prices, the sellers update the amount of electricity and the game continues until convergence to a Stackelberg equilibrium. In \cite{P2PEnergyTradingProsumerCommMicrogrid2019Paudeletal}, the strategies are quite different. The sellers announce their prices and the amount of electricity they are able to sell first, and then the buyers react with a selection of the sellers. This selection is a probability distribution for each buyer over the complete set of sellers and should indicate the probability of a buyer choosing a specific seller. The buyers compute this selection using an evolutionary game. Based on this selection, the sellers update their prices using a non-cooperative game. Note that hereby the amount of electricity to sell is a fixed parameter in this setting. Again, it is shown that the iterative Stackelberg game converges to a Stackelberg equilibrium. In \cite{LETPriceUncertStackelGame2019ElRahietal}, \cite{EnergySharingStochProgStackGame2018Liuetal} and \cite{P2PpricebasedDRMicrogrids2017Liuetal} on the other hand, the prosumers are not in advance divided into buyers or sellers. In all these settings, the leader is a single entity that can buy and sell electricity and the followers are the set of prosumers. In \cite{EnergySharingStochProgStackGame2018Liuetal} and \cite{P2PpricebasedDRMicrogrids2017Liuetal}, the leader is a storage system within the microgrid, which can buy excess electricity or sell electricity to prosumers with a demand. Its goal is to maximize its profit, while the objectives of the prosumers are to maximize their own utility. The leader starts by announcing internal prices for the prosumers. Based on these prices, the prosumers can each solve their own (bounded) optimization problem to maximize their utility. They then announce their optimal amount of electricity to buy or sell, and the leader reacts to this by adjusting its prices. While the convergence of this iterative approach to a Stackelberg game is shown in \cite{EnergySharingStochProgStackGame2018Liuetal}, in \cite{P2PpricebasedDRMicrogrids2017Liuetal} the model is based on a bi-level optimization problem, and no guarantees for convergence are made. In \cite{LETPriceUncertStackelGame2019ElRahietal}, instead of a storage system, a power company is the leader of the Stackelberg game. It first announces a price, based on which the prosumers play a non-cooperative game among themselves to determine how much electricity to buy or sell. Two different ways to achieve a Stackelberg game are proposed, with the first one being an iterative one leading to an $\epsilon$-Stackelberg equilibrium, while in the second, the leader solves a non-linear optimization problem to directly find the Stackelberg equilibrium.\\

The second group of settings uses Stackelberg games to model the relation between the prosumers and their MGO or DSO. This setting is considered in \cite{LocalTradingFramework2020Askeland}, \cite{TrilevelBestResponseDemandSide2020Ausseletal},  \cite{EnergySharingDistrGamePricing2018CuiWangLiu}, \cite{LEMsEfficiencyMultiLeaderSG2019LeCadre}, \cite{AGTcoalitionFairStablePricing2018LeCadreetal},  \cite{AGTEnergySharingPVStackelbergGame2017Liuetal},  \cite{AGTCollaborativeEnergyManagNonCoopEvolApproach2019Rajasekharetal}, \cite{AGTPrioritizeConsumers2014Tushar} and \cite{bilevelDemandResponseMarket2013Zugnoetal}. In \cite{EnergySharingDistrGamePricing2018CuiWangLiu}, \cite{AGTcoalitionFairStablePricing2018LeCadreetal} and \cite{AGTEnergySharingPVStackelbergGame2017Liuetal}, the leader of the game is the MGO, while the prosumers are the followers. The goal of the leader is to maximize its profit and it starts by submitting initial internal buying and selling prices to the prosumers. The prosumers use these prices as input to their utility maximization problems and optimize them on their own. The prosumers then announce the amount of electricity to buy or sell and the leader updates its prices. The existence of a Stackelberg equilibrium is shown. In \cite{LocalTradingFramework2020Askeland}, \cite{LEMsEfficiencyMultiLeaderSG2019LeCadre},  \cite{AGTPrioritizeConsumers2014Tushar} and \cite{bilevelDemandResponseMarket2013Zugnoetal}, the leader is either a central power station, which wants to buy surplus electricity from the prosumers, the DSO, which wants to minimize the grid cost of the microgrid or retailers, who want to maximize their profit of selling electricity to the prosumers. The followers are once again the prosumers, who want to maximize their utility, or local MGOs, who want to maximize the social welfare of their set of prosumers. The leader announces initial prices or grid tariffs and based on this, the prosumers optimize their utility. In contrast to \cite{EnergySharingDistrGamePricing2018CuiWangLiu} and \cite{AGTEnergySharingPVStackelbergGame2017Liuetal}, the prosumers either solve a generalized Nash equilibrium (GNE) game to decide how much to sell to the central power station, or they need to solve a complementary problem to compute an equilibrium. In both cases, the prosumers then announce their electricity consumption, either on an individual base, \cite{AGTPrioritizeConsumers2014Tushar}, or on an aggregated level \cite{LocalTradingFramework2020Askeland}. Based on the reaction of the prosumers, the leader updates its prices and this iterative scheme continues until some convergence criterion is met. In \cite{bilevelDemandResponseMarket2013Zugnoetal} on the other hand, the bilevel problem is reformulated into a single-level MILP, which can easily be solved. A similar approach is taken in \cite{TrilevelBestResponseDemandSide2020Ausseletal}, where the trilevel problem is reformulated twice to obtain a tractable formulation. In the first step, an explicit formulation is derived for the prosumers, which removes the bottom layer. The remaining two layers, with the supplier being the leader and the MGOs being followers, are then reformulated using the KKT conditions of the followers in the leader's problem. In \cite{AGTCollaborativeEnergyManagNonCoopEvolApproach2019Rajasekharetal}, the MGO also acts as the leader of the Stackelberg game, but instead of using price signals as a strategy, it uses demand profiles. In the beginning, the MGO collects the load profiles of all prosumers and optimally schedules its own battery usage. It then broadcasts the aggregated load profiles minus the prosumer's load profile to each prosumer. In addition, also boundaries for the aggregated load profile and penalty prices are announced. The prosumers then optimize their utility function, which is a weighted sum of electricity costs, the comfort level, and the minimization of interruption to increase the life span of appliances. The prosumers announce their resulting load profiles to the MGO, which updates its battery schedule and possibly also the penalty prices. This process continues until the difference in the objective function of the MGO is reasonably small. It is shown that the iterative process converges to a Stackelberg equilibrium.\\

The third group of approaches uses general non-cooperative games to model the interactions between prosumers and other participants. The corresponding approaches are \cite{LoadSheddingRHstochCompEqui2018DevineBertsch}, \cite{MarketEquilibriumInfoAsymmetry2019Dvorkinetal},  \cite{MarketEquilibriaStorageADMM2020Gruebeletal},  \cite{bidirectionalTradingSchedulingEV2013Kimetal}, \cite{P2PMarketAanlysisVariationaltoGNE2020LeCadreetal}, \cite{MultiMicrogridDSM2021Li},  \cite{GNEanalysisP2PMarketDistGrid2021ShilovLeCadreBuvsic}, \cite{AGTPriceDiscriminationEnergyTrading2017Tusharetal} and \cite{P2PEnergyTradingDR2019Zhangetal}. In \cite{LoadSheddingRHstochCompEqui2018DevineBertsch}, \cite{MarketEquilibriaStorageADMM2020Gruebeletal}, \cite{bidirectionalTradingSchedulingEV2013Kimetal} and \cite{P2PMarketAanlysisVariationaltoGNE2020LeCadreetal}, a non-cooperative game is played among all prosumers. The utility functions of the players consist of the cost and the satisfaction of electricity consumption, while the strategies of the prosumers are their load profiles. In \cite{MarketEquilibriaStorageADMM2020Gruebeletal}, the prosumer may be equipped with storage devices and the market equilibrium problem is reformulated into a mixed complementarity problem using the KKT conditions. The resulting formulation is then solved via ADMM. A similar solution approach is presented in \cite{LoadSheddingRHstochCompEqui2018DevineBertsch}, where a non-cooperative game between prosumers, consumers, and generator units with possible failure times is modeled. The formulation results in a stochastic mixed complementarity problem, which solves the optimization problems of each prosumer and results in an equilibrium solution. In \cite{bidirectionalTradingSchedulingEV2013Kimetal}, a tailor-made billing scheme penalizes heavy electricity users, and it is shown that an iterative gradient-based algorithm converges to the NE of the game. In \cite{P2PMarketAanlysisVariationaltoGNE2020LeCadreetal} on the other hand, the coupling constraints between the prosumers lead to a GNE. A detailed analysis provides insights into the efficiency of the GNE compared to a central solution. In \cite{GNEanalysisP2PMarketDistGrid2021ShilovLeCadreBuvsic}, a GNE game is played between the DSO and the prosumers. The strategy of the prosumers is based on the amount of flexibility that they are willing to offer, while the strategy of the DSO is based on the fraction of the prosumers' flexibility that it wants to use, as well as a congestion price. Due to a coupling constraint between the prosumers as well as the DSO, a GNE is computed. In \cite{AGTPriceDiscriminationEnergyTrading2017Tusharetal}, a similar setting is considered, in which a power company is interested in buying surplus electricity from the prosumers, given a fixed budget. A cake-cutting game is proposed and a variational equilibrium is found using a decentralized algorithm. In \cite{MarketEquilibriumInfoAsymmetry2019Dvorkinetal}, a non-cooperative game between a market operator, producers, and consumers is modeled. It is shown that a unique equilibrium always exists and a distributed algorithm is presented, in which producers and consumers react to the market operator's prices by adapting their production and consumption. In \cite{P2PEnergyTradingDR2019Zhangetal}, two non-cooperative games between prosumers, the MGO, and suppliers are played. The MGO acts as a local aggregator between the prosumers on the one side and the suppliers on the other side. For the non-cooperative game between MGO and suppliers, the suppliers offer bids to the MGO. The MGO then uses these bids and the net demand of the prosumers to compute external trading prices with the suppliers. The utility function of the suppliers represents the profit they make by selling electricity to the MGO. The second non-cooperative game in \cite{P2PEnergyTradingDR2019Zhangetal} is played among the prosumers, who decide on their load profiles, given some predefined buying and selling prices for the given time interval. The utility functions of the prosumers consist of the cost of buying or the profit of selling electricity locally as well as the utility of electricity consumption. For both non-cooperative games, it is shown that a unique NE exists and an iterative algorithm is given, which converges to the NE. Both non-cooperative games are then connected via the MGO, which updates the external and internal prices after a change in either bids from the suppliers or the electricity consumption from the prosumers. In \cite{MultiMicrogridDSM2021Li} on the other hand, a trilevel problem between the DSO, MGOs, and prosumers is modeled. Two different solution approaches, one cooperative and one non-cooperative are proposed. In both cases, the trilevel model is first reduced to a bilevel model by deriving an analytical solution to the non-cooperative game between MGOs and their respective prosumers. The remaining bilevel problem is then solved either in a cooperative or non-cooperative way using price and demand as signals.\\

The last group combines Stackelberg or general non-cooperative games with auctions, see \cite{P2PETDoubleAuction2021DoanChoKim}, \cite{DAadaptiveSharingMarket2021HeZhang}, \cite{AGTnoncoopDoubleAuctionPHEVTrading2011Saadetal}, \cite{AGTtransactiveEnergyFlexiblePros2021Georgiosetal}, \cite{AGTAuctionBasedStorageSharing2016Tusharetal} and \cite{GameTheoryEnergyTradingSG2014Wangetal}. In the considered literature, there are two main ways to combine these approaches with each other. In \cite{AGTnoncoopDoubleAuctionPHEVTrading2011Saadetal} and \cite{GameTheoryEnergyTradingSG2014Wangetal}, a non-cooperative game is played among a set of prosumers with a surplus of electricity. The strategies of the sellers are specified by the amount of electricity they are willing to sell, while the utility is the profit they gain by selling electricity to the buyers. The prices are computed using a standard double auction between buyers and sellers, as is also often seen in the Strategic Matching without Flexibility cluster. After initializing the amounts to sell, the double auction is run, and based on the new clearing price, each prosumer one after the other finds best responses by communicating with the MGO, which acts as the auctioneer. It is shown that this iterative algorithm converges to a NE. In \cite{DAadaptiveSharingMarket2021HeZhang}, a non-cooperative game is played among the prosumers of a microgrid. Each prosumer first solves a simple optimization problem to determine how much electricity to offer or ask for in the auction. Following a double auction, the winners participate in a non-cooperative game, in which each participant finds an optimal deviation from its original bid. This deviation maximizes a utility function, which consists of profit and the reluctance to deviate from the original bid. Afterward, the clearing price of the double auction is updated and the non-cooperative game continues, with each participant finding its best response to the new clearing price until a stopping criterion is met. Similarly to the above approaches, in \cite{AGTtransactiveEnergyFlexiblePros2021Georgiosetal}, a modified version of a combinatorial auction is run. Within each iteration, players are added to the set of winners of the auction, based on the outcome of a non-cooperative game. In \cite{P2PETDoubleAuction2021DoanChoKim} and \cite{AGTAuctionBasedStorageSharing2016Tusharetal} on the other hand, a double auction is run first to determine the set of winners of the auction, as well as the clearing price limits. Then, a Stackelberg game is played, with the MGO being the leader and the followers are either the winning buyers or sellers of the double auction. Using the range of possible clearing prices, the objective of the MGO is to maximize the average social welfare of the remaining set of winners. The strategy of the followers is to adapt the amount of electricity they are willing to sell or buy. This iterative process continues until the Stackelberg equilibrium is found.

\subsection{Connections between Clusters}\label{SSBetweenClusters}
Based on the previous analysis of the clusters w.r.t. the used techniques and various aspects of the settings, we now identify and highlight connections between the three main clusters. Thereby, approaches from different clusters may complement each other when combined.\\

As already seen in the Equilibrium Balancing with Flexibility cluster, there are different approaches that combine the flexibility of this cluster with the auction-based approaches in the Matching without Flexibility cluster. This allows for an integration of a market-based pricing scheme into prosumers' decision processes and thereby extends the given approaches. \\

Another possible combination of approaches from two different clusters is to use the decentralized algorithms presented in the Joint Balancing with Flexibility cluster to compute an optimal solution, which can then be used in the Cooperative Pricing Scheme cluster. In both cases, the objective of the optimization problem is to minimize the sum of electricity costs of the microgrid. While the approaches in the Cooperative Pricing Scheme cluster are often based on centralized approaches along with their disadvantages regarding data privacy, decentralized optimization approaches (in the Joint Balancing with Flexibility cluster) could avoid this. In addition, new possibilities on how to fairly assign the benefits of cooperating among the participants may arise from this connection.

\section{Discussion and Conclusion}\label{SConclusion}

Summarizing, we can state that there are currently three main lines of research for local energy trading, each with a distinct setting and focus on one specific high-level research question of local energy trading:
\begin{enumerate}
	\item The main goal of approaches within the Matching without Flexibility cluster is to match demand and supply, mostly by means of an auction. Prosumers can express their individual valuation in the form of bids and the MGO computes a clearing price, which maximizes the social welfare using well-established auction mechanisms.
	\item The Communal Pricing Scheme focuses on creating pricing mechanisms that incentivize prosumers to be part of a microgrid. Instead of focusing on the load and flexibility of single prosumers, load profiles and flexibility of the whole community are combined to increase the overall profit. This additional profit is then split up between the prosumers, such that everyone profits from participation.
	\item The last cluster is the Balancing with Flexibility cluster, which makes use of the flexibility of smart devices and the present loads to balance demand and supply. Using decentralized algorithms on the base of the underlying structure of the microgrid, the privacy of data can be ensured.
\end{enumerate}
These central research questions within the clusters also align well with the considered research questions and findings of the analyzed literature. Combined with the different characteristics of the classification scheme, see \Cref{SClustering}, we can identify several future research directions and major open problems.\\

The first research direction is the uncertainty in data. While it has already been considered in other related areas, such as power flow computations, this topic has not been studied much in a local energy trading setting. Most of the considered trading approaches simply use predictions or forecasts for the majority of their data. Although many publications across all clusters state the integration of uncertainty or stochasticity in data as an important, open research question, only a few approaches have already done first steps to integrate uncertainty into their methods. We propose to start with the integration of robust or stochastic methods into approaches of the Joint Balancing with Flexibility cluster and then to further proceed with approaches in other clusters. This is based on the following reasons: 
\begin{enumerate}[label=1.\arabic*]
	\item The original formulation of these approaches are all built upon a centralized model, for which it is fairly straightforward to apply well-established methods, such as (adaptive) robust optimization or stochastic programming. The challenge then lies within the adaption of the decentralized optimization techniques to the updated formulations of the problem.
	\item Most of the settings within the Joint Balancing with Flexibility cluster can cover time horizons of several hours and more. Within these time horizons, a considerable amount of uncertainty, compared to short time horizons of up to 15 min, can appear. Therefore, it is important to deal with the uncertainty in these settings first, before also considering smaller time horizons.
	\item When applying techniques from robust optimization to the centralized models in the Joint Balancing with Flexibility cluster, additional synergy effects, due to uncertainty sets, may appear, when combining the robust approaches with the ideas of the Cooperative Pricing Scheme cluster.
\end{enumerate}
It is worth mentioning that the integration of uncertainty has to be applied not only for local energy trading but also for energy management approaches, which are similar to the approaches in Joint Balancing with Flexibility.\\

The second future research direction, which up to now has not received much attention in local energy trading, is the integration of the physical infrastructure of microgrids. In large parts of the analyzed literature, the power aspect of the (LV) grid is strictly separated from the energy aspect of local trading. Future work in local energy trading has to integrate the power aspect into the approaches to ensure a stable electricity distribution. There are two main ways how to achieve this:
\begin{enumerate}[label=2.\arabic*]
	\item The direct approach integrates the power aspect by means of constraints, directly modeling the power flow and grid constraints. Approaches in the Joint Balancing with Flexibility cluster are once again a good starting point due to their central optimization models, in which different versions of power flow approximations can easily be integrated.
	\item The indirect approach makes use of different steering approaches, such as e.g. congestion prices. These can then encourage or discourage prosumers to consume more electricity or to curtail their PV generation. Approaches within the Matching without Flexibility clusters are a good starting point to integrate the power aspect in an indirect way, see e.g. \cite{TEAuctionTradingActiveProsumer2017KhorasanyMishraLedwich}, \cite{TradingAgentSmartGrid2010Vytelingumetal}.
\end{enumerate}

A third research direction, which has yet to gain focus in local energy trading, is the modeling of human behavior, see e.g. \cite{P2PTradingEnergyCommunityModelling2022PenaBelloetal}. While it is reasonable (and necessary) to make some assumptions on prosumer behavior to analyze equilibria and their efficiencies, it has been shown that prosumers do not focus solely on the financial aspect of their decisions \cite{AffectingConsumerBehavior2007Martiskainen}. Based on the reviewed literature, two ways how to integrate prosumer behavior into local energy trading can be identified:
\begin{enumerate}[label=3.\arabic*]
	\item A rather direct approach is to introduce different classifications of electricity, representing various aspects, such as 'green' or 'local' electricity. Prosumers can then follow their individual preferences over these different types, see e.g. \cite{P2PEnergyTradingMultiClassEM2018MorstynMcculloch}. 
	\item Another approach is to extend the already existing utility functions of the prosumers by additional aspects, such as e.g. an ecological motive. Using game-theoretic approaches, a detailed analysis of equilibria w.r.t. differently weighted motives could reveal interesting results and insights for designing future energy policies and incentives.
\end{enumerate}

\section*{Acknowledgements}
	This research is supported by the Netherlands Organization for Scientific Research (NWO) grant 645.002.001.

\section*{Authors' contributions}
All authors contributed to the design of the project and to the literature search. Jens H\"onen took the lead in writing the first draft of the paper, all authors contributed to realizing the final version.

\printbibliography

\end{document}